\documentclass[12pt]{iopart}
\newtheorem{theorem}{Theorem}[section]

\newtheorem{lemma}[theorem]{Lemma}

\eqnobysec
\begin{document}
\jl{1}
\title[Analytic Bethe ansatz]{Analytic Bethe ansatz and functional equations 
 for Lie superalgebra $sl(r+1|s+1)$
}
\author{Zengo Tsuboi}

\address{Institute of Physics,  
Graduate School of Arts and Sciences \\                                 
 University of Tokyo, Komaba \\
  3-8-1 Komaba, Meguro-ku, Tokyo 153, JAPAN}

\begin{abstract}
From the point of view of the Young superdiagrm method,
 an analytic Bethe ansatz is carried out for Lie 
superalgebra $sl(r+1|s+1)$. For the transfer matrix 
eigenvalue formulae in dressed vacuum form,
 we present some expressions,  
which are quantum analogue of Jacobi-Trudi and 
Giambelli formulae for Lie superalgebra $sl(r+1|s+1)$. 
We also propose transfer matrix functional relations,
 which are Hirota bilinear difference equation with some constraints.   
\end{abstract}

\noindent Report number: UT-Komaba/97-8 \\
Journal-ref: J. Phys. A: Math. Gen. 30 (1997) 7975-7991 \\ 
DOI: 10.1088/0305-4470/30/22/031

\maketitle

\section{Introduction}
In [KNS1], a class of functional relations, the T-system, was proposed. 
It is a family of functional relations for a set of commuting transfer 
matrices of solvable lattice models associated with any quantum 
affine algebras $U_q({\cal G}_{r}^{(1)})$. Using T-system, we can
 calculate various physical quantities [KNS2] such as the 
 correlation lengths of the vertex models and central charges
 of RSOS models. The T-system is not only a family of 
 transfer matrix functional 
 relations but also two-dimensional Toda field equation 
on discrete space time. And it has beautiful pfaffian and determinant 
solutions [KOS,KNH,TK] (see also, [T]).
 
In [KS1], analytic Bethe ansatz [R1] was carried out for 
fundamental representations of the Yangians $Y({\cal G})$[D], where 
${\cal G}=B_{r}$, $C_{r}$ and $D_{r}$. That is, eigenvalue
 formulas in dressed vacuum form were proposed for the 
 transfer matrices of solvable vertex models .
  These formulae are Yangian analogues of the Young tableaux 
 for ${\cal G}$ and satisfy certain semi-standard like 
conditions. It had been proven that they are free of poles under the 
Bethe ansatz equation. Furthermore, for ${\cal G}=B_{r}$ case, these
formulae were extended to the case of finite dimensional modules 
labeled by skew-Young diagrams $\lambda \subset \mu$ [KOS].
 In analytic Bethe ansatz context, 
above-mentioned solutions of the T-system correspond 
to the eigenvalue formulae of the transfer matrices in dressed vacuum form 
labeled by rectangular-Young diagrams 
$\lambda =\phi, \mu=(m^a)$ (see also, [BR,KLWZ,K,KS2,S2]).

The purpose of this paper is to extend similar 
analyses to Lie superalgebra ${\cal G}=sl(r+1|s+1)$ [Ka] case 
(see also [C] for comprehensible account on Lie superalgebras).
Throughout this paper,
 we frequently use similar
 notation presented in [KS1,KOS,TK].
Studying supersymmetric integrable models is important not only 
 in mathematical physics but also in condensed matter physics
 (cf.[EK,FK,KE,S1,ZB]). 
 For example, the supersymmetric $t-J$ model received much 
 attention in connection with  high $T_{c}$ superconductivity.  
In the supersymmetric models, the R-matrix satisfies the graded
Yang-Baxter equation [KulSk]. The transfer matrix is defined as a 
 {\it super} trace of monodromy matrix. 
 As a result, extra signs appear in the Bethe ansatz equation and 
 eigenvalue formula of the transfer matrix.

There are several inequivalent choices of simple root system 
 for Lie superalgebra. We treat so-called distinguished simple 
 root system [Ka] in the main text. 
We introduce the Young superdiagram [BB1], which is associated with a 
 covariant tensor representation. 
 To be exact, this Young superdiagram is different from the classical
  one  in that it carries spectral parameter $u$. 
 In contrast to ordinary Young diagram, 
there is no restriction on the number of rows. We define 
 semi-standard like tableau on it. 
 Using this tableau, we introduce the function 
${\cal T}_{\lambda \subset \mu}(u)$ (\ref{Tge1}).
 This should be fusion transfer 
matrix of dressed vacuum form in the analytic Bethe ansatz.
We prove pole-freeness of ${\cal T}^{a}(u)={\cal T}_{(1^{a})}(u)$, 
crucial property of analytic Bethe ansatz.
Due to the same mechanism presented in [KOS], the function
 ${\cal T}_{\lambda \subset \mu}(u)$ has 
determinant expression whose matrix elements are only 
the functions associated with Young superdiagrams with shape 
$\lambda = \phi $; $\mu =(m)$ or $(1^{a})$. 
It can be viewed as quantum analogue of Jacobi-Trudi and 
Giambelli formulae for Lie superalgebra $sl(r+1|s+1)$.  
Then one can easily 
show that the function ${\cal T}_{\lambda \subset \mu}(u)$ 
is free of poles under 
the Bethe ansatz equation (\ref{BAE}).
 Among the above-mentioned eigenvalue formulae of 
transfer matrix in dressed vacuum form 
associated with rectangular Young superdiagrams, 
we present a class of transfer matrix functional relations. 
It is a special case of Hirota bilinear difference equation [H]. 

Deguchi and Martin [DM] discussed the spectrum of fusion model 
from the point of view of representation theory (see also, [MR]).
The present paper will partially give us elemental account on their 
result from the point of view of the analytic Bethe ansatz.

The outline of this paper is given as follows.
In section2, we execute analytic Bethe ansatz 
based upon the Bethe ansatz equation (\ref{BAE}) associated 
with the distinguished simple roots. 
The observation that the Bathe ansatz equation can be expressed 
by root system of Lie algebra is traced back to [RW]
 (see also, [Kul] for $sl(r+1|s+1)$ case). 
 Moreover, Kuniba et.al.[KOS] conjectured  that left hand side of the 
 Bethe ansatz equation (\ref{BAE}) can be written as a ratio 
  of certain \symbol{96}Drinfeld polynomials' [D].  
We introduce the function ${\cal T}_{\lambda \subset \mu}(u)$,
 which should be the transfer 
matrix whose auxiliary space is finite dimensional 
 module of super Yangian $Y(sl(r+1|s+1))$ [N] or 
quantum affine superalgebra $U_{q}(sl(r+1|s+1)^{(1)})$ [Y],  
labeled by skew-Young superdiagram $\lambda \subset \mu$.
The origin of the function ${\cal T}^{1}(u)$ goes back to the 
 eigenvalue formula of 
transfer matrix of Perk-Schultz model [PS1,PS2,Sc], 
which is a multi-component generalization of the 
six-vertex model (see also [Kul]).  
In addition, the function ${\cal T}^{1}(u)$ reduces to 
the eigenvalue formula of transfer matrix derived by 
algebraic Bethe ansatz (For example, [FK]:$r=1,s=0$ case; 
[EK]:$r=0,s=1$ case; [EKS1,EKS2]:$r=s=1$ case).
In section3, we propose  functional relations,
 the T-system, associated with the transfer matrices 
 in dressed vacuum form defined in the previous section.
Section4 is devoted to summary and discussion. 
In appendix A, we briefly mentioned relation between 
fundamental $L$ operator and the transfer matrix.
In this paper,  we treat mainly the expressions 
related to covariant representations.  
For contravariant ones, we 
 present several expressions in Appendix B.  
 Appendix C and D provide some expressions related to  
 non-distinguished simple roots of $sl(1|2)$. 
 Appendix E explains how to represent the eigenvalue formulae of 
transfer matrices in dressed vacuum form
${\cal T}_{m}(u)$ and ${\cal T}^{a}(u)$   
 in terms of the functions 
${\cal A}_{m}(u)$, ${\cal A}^{a}(u)$,
${\cal B}_{m}(u)$ and ${\cal B}^{a}(u)$, 
which are analogous to the fusion 
transfer matrices of $U_q({\cal G}^{(1)})$ 
vertex models (${\cal G}=sl_{r+1}, sl_{s+1}$).
\section{Analytic Bethe ansatz}
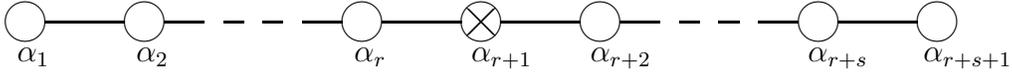
\begin{figure}
    \setlength{\unitlength}{0.75pt}
    \begin{center}
    \begin{picture}(480,30) 
      \put(10,20){\circle{20}}
      \put(20,20){\line(1,0){40}}
      \put(70,20){\circle{20}}
      \put(80,20){\line(1,0){20}}
      \put(110,20){\line(1,0){10}}
      \put(130,20){\line(1,0){10}}
      \put(150,20){\line(1,0){20}}
      \put(180,20){\circle{20}}
      \put(190,20){\line(1,0){40}}
      \put(240,20){\circle{20}}
      \put(232.929,12.9289){\line(1,1){14.14214}}
      \put(232.929,27.07107){\line(1,-1){14.14214}}
      \put(250,20){\line(1,0){40}}
      \put(300,20){\circle{20}}
      \put(310,20){\line(1,0){20}}
      \put(340,20){\line(1,0){10}}
      \put(360,20){\line(1,0){10}}
      \put(380,20){\line(1,0){20}}
      \put(410,20){\circle{20}}
      \put(420,20){\line(1,0){40}}
      \put(470,20){\circle{20}}
      \put(6,0){$\alpha_{1}$}
      \put(66,0){$\alpha_{2}$}
      \put(176,0){$\alpha_{r}$}
      \put(235,0){$\alpha_{r+1}$}
      \put(295,0){$\alpha_{r+2}$}
      \put(405,0){$\alpha_{r+s}$}
      \put(463,0){$\alpha_{r+s+1}$}
  \end{picture}
  \end{center}
  \caption{Dynkin diagram for the Lie superalgebra
  $sl(r+1|s+1)$ corresponding to the distinguished simple roots:
   white circle denotes even root $\alpha_i$; 
   grey (a cross) circle denotes odd root $\alpha_j$ with 
   $(\alpha_j|\alpha_j)=0$.}
   \label{distinguished}
\end{figure}

Lie superalgebra [Ka] is a ${\bf Z}_2$ graded algebra 
${\cal G} ={\cal G}_{\bar{0}} \oplus {\cal G}_{\bar{1}}$ 
with a product $[\; , \; ]$, whose homogeneous
elements $a\in {\cal G_{\alpha}},b\in {\cal G_{\beta}}$ 
$(\alpha, \beta \in {\bf Z}_2=\{\bar{0},\bar{1} \})$ and
 $c\in {\cal G}$ satisfy the following relations.
\begin{eqnarray}
\left[a,b\right] \in {\cal G}_{\alpha+\beta}, \nonumber \\ 
\left[a,b\right]=-(-1)^{\alpha \beta}[b,a], \\
\left[a,[b,c]\right]=[[a,b],c]+(-1)^{\alpha \beta} [b,[a,c]].
 \nonumber  
\end{eqnarray}
The set of non-zero roots can be divided into the set of non-zero  
even roots (bosonic roots) $\Delta_0^{\prime}$ and the set of odd roots 
(fermionic roots) $\Delta_1$. For $sl(r+1|s+1)$ case, they read 
\begin{equation}
\Delta_0^{\prime}=
 \{ \epsilon_{i}-\epsilon_{j} \} \cup \{\delta_{i}-\delta_{j}\}, 
i \ne j ;\quad \Delta_1=\{\pm (\epsilon_{i}-\delta_{j})\}
\end{equation}
where $\epsilon_{1},\dots,\epsilon_{r+1};\delta_{1},\dots,\delta_{s+1}$ 
are basis of dual space of the Cartan subalgebra with the bilinear 
form $(\ |\ )$ such that 
\begin{equation}
 (\epsilon_{i}|\epsilon_{j})=\delta_{i\, j}, 
 (\epsilon_{i}|\delta_{j})=(\delta_{i}|\epsilon_{j})=0 ,
 (\delta_{i}|\delta_{j})=-\delta_{i\, j}. 
\end{equation} 
There are several choices of simple root system reflecting choices 
of Borel subalgebra. 
The simplest system of simple roots is so called distinguished one [Ka] 
(see, figure \ref{distinguished}).
Let $\{\alpha_1,\dots,\alpha_{r+s+1} \}$ be the distinguished simple 
roots of Lie superalgebra $sl(r+1|s+1)$ 
 \begin{eqnarray}
   \alpha_i = \epsilon_{i}-\epsilon_{i+1}
    \quad i=1,2,\dots,r, \nonumber \\
   \alpha_{r+1} = \epsilon_{r+1}-\delta_{1}  \\ 
    \alpha_{j+r+1} = \delta_{j}-\delta_{j+1} ,
    \quad j=1,2,\dots,s \nonumber   
 \end{eqnarray}
 and with the grading 
\begin{equation}
    {\rm deg}(\alpha_a)=\left\{
              \begin{array}{@{\,}ll}
                0  & \mbox{for even root} \\ 
                1 & \mbox{for odd root} 
              \end{array}
            \right. 
\end{equation}
Especially for distinguished simple root, we have 
$\deg(\alpha_{a})=\delta_{a,r+1}$.

We consider the following type of the Bethe
 ansatz equation (cf. [Kul,RW,KOS]).
\numparts
\begin{eqnarray}
-\frac{P_{a}(u_k^{(a)}+\frac{1}{t_{a}})}{P_{a}(u_k^{(a)}-\frac{1}{t_{a}})}
   =(-1)^{\deg(\alpha_a)} 
    \prod_{b=1}^{r+s+1}\frac{Q_{b}(u_k^{(a)}+(\alpha_a|\alpha_b))}
           {Q_{b}(u_k^{(a)}-(\alpha_a|\alpha_b))}, \label{BAE} \\ 
        Q_{a}(u)=\prod_{j=1}^{N_{a}}[u-u_j^{(a)}],
        \label{Q_a} \\ 
        P_{a}(u)=\prod_{j=1}^{N}P_{a}^{(j)}(u), \\  
        P_{a}^{(j)}(u)=[u-w_{j}]^{\delta_{a,1}} \label{drinfeld},
\end{eqnarray}
\endnumparts
where $[u]=(q^u-q^{-u})/(q-q^{-1})$; $N_{a} \in {\bf Z }_{\ge 0}$; 
$u, w_{j}\in {\bf C}$; $a,k \in {\bf Z}$ 
($1\le a \le r+s+1$,$\ 1\le k\le N_{a}$);
 $t_{a}=1$ for $1\le a \le r+1$,
 $t_{a}=-1$ for $r+2\le a \le r+s+1$.
In this paper, we suppose that $q$ is generic.
The left hand side of the Bethe ansatz equation (\ref{BAE}) 
is related to the quantum space.
We suppose that it is given by the ratio of some 
\symbol{96}Drinfeld polynomials'  labeled by skew-Young diagrams
  $\tilde{\lambda} \subset \tilde{\mu}$ (cf.[KOS]).
  For simplicity, we consider only the case 
   $\tilde{\lambda}=\phi, \tilde{\mu}=(1) $.
The generalization to the case for any skew-Young diagram
will be achieved by the empirical procedures mentioned in [KOS].
 The factor $(-1)^{\deg(\alpha_a)}$ of 
the Bethe ansatz equation (\ref{BAE}) appears so as to make
 the transfer matrix to be a {\it super} 
trace of monodromy matrix.
We define the sets 
\begin{eqnarray}
      J=\{ 1,2,\dots,r+s+2\} ,
     \quad    J_{+}=\{ 1,2,\dots,r+1\} , \nonumber \\ 
        J_{-}=\{ r+2,r+3,\dots,r+s+2\},  
  \label{set}
\end{eqnarray}
with the total order 
\begin{eqnarray} 
 1\prec 2 \prec \cdots \prec r+s+2 \label{order}
\end{eqnarray}
and with the grading 
\begin{equation}
      p(a)=\left\{
              \begin{array}{@{\,}ll}
                0  & \mbox{for $a \in J_{+}$}  \\ 
                1 & \mbox{for $a \in J_{-}$ }
                 \quad .
              \end{array}
            \right. \label{grading}
\end{equation}
For $a \in J $, set
\begin{eqnarray}
\fl   z(a;u)=\psi_{a}(u)
     \frac{Q_{a-1}(u+a+1)Q_{a}(u+a-2)}{Q_{a-1}(u+a-1)Q_{a}(u+a)} 
    \qquad {\rm for} \quad  a \in J_{+},
        \nonumber \\
\fl     z(a;u)=\psi_{a}(u) \frac{Q_{a-1}(u+2r-a+1)Q_{a}(u+2r-a+4)}
        {Q_{a-1}(u+2r-a+3)Q_{a}(u+2r-a+2)} 
    \qquad {\rm for} \quad  a \in J_{-},
    \label{z+}
\end{eqnarray}
where $Q_{0}(u)=1, Q_{r+s+2}(u)=1$ and 
\begin{equation}
  \psi_{a}(u)=
   \left\{
    \begin{array}{@{\,}ll}
      P_{1}(u+2) & \mbox{for } \quad a=1 \\ 
      P_{1}(u) & \mbox{for } \quad a \in J-\{1\}
    \end{array} \label{psi}
   \right. .
\end{equation}
In this paper, we often express the function $z(a;u)$ as the box
 $\framebox{a}_{u}$, whose spectral parameter $u$ will often 
 be abbreviated. Under the Bethe ansatz equation,
 we have 
\begin{numparts}
\begin{eqnarray}
\fl Res_{u=-b+u_{k}^{(b)}}(z(b;u)+z(b+1;u))=0 
    \quad 1\le b \le r \label{res1} \\
\fl Res_{u=-r-1+u_{k}^{(r+1)}}(z(r+1;u)-z(r+2;u))=0  
     \label{res2} \\
\fl Res_{u=-2r-2+b+u_{k}^{(b)}}(z(b;u)+z(b+1;u))=0 
     \quad r+2\le b \le r+s+1 \label{res3}
\end{eqnarray}
\end{numparts}
We will use the functions ${\cal T}^{a}(u)$ 
and ${\cal T}_{m}(u)$ 
 ($a \in {\bf Z }$; $m \in {\bf Z }$; $u \in {\bf C }$) 
 determined by the following generating series
\begin{numparts} 
\begin{eqnarray}
\fl     (1+z(r+s+2;u)X)^{-1}\cdots (1+z(r+2;u)X)^{-1}
     (1+z(r+1;u)X)\cdots (1+z(1;u)X) \nonumber \\
\fl     =\sum_{a=-\infty}^{\infty} 
         {\cal F}^{a}(u+a-1)
         {\cal T}^{a}(u+a-1)X^{a},
         \label{generating}
\end{eqnarray} 
\begin{equation}
\fl {\cal F}^{a}(u)= 
 \left\{
    \begin{array}{@{\,}ll}
      \prod_{j=1}^{a-1} P_{1}(u-2j+a-1) & \mbox{for } \quad a \ge 2 \\ 
      1                    & \mbox{for } \quad a=1 \\ 
      \frac{1}{P_{1}(u-1)} & \mbox{for } \quad a=0 \\
      0                    & \mbox{for } \quad a \le -1 \\  
    \end{array}
   \right. ,
\end{equation}
\begin{eqnarray}
\fl     (1-z(1;u)X)^{-1}\cdots (1-z(r+1;u)X)^{-1}
      (1-z(r+2;u)X)\cdots (1-z(r+s+2;u)X)\nonumber \\
\fl     =\sum_{m=-\infty}^{\infty} {\cal T}_{m}(u+m-1)X^{m}, 
\end{eqnarray}
\end{numparts} 
where $X$ is a shift operator $X=\e^{2\partial_{u}}$. 
In particular, we have ${\cal T}^{0}(u)=P_{1}(u-1)$; 
${\cal T}_{0}(u)=1$; ${\cal T}^{a}(u)=0$ for $a<0$ ; 
 ${\cal T}_{m}(u)=0$ for $m<0$. 
We remark that the origin of the function ${\cal T}^{1}(u)$ and 
the Bethe ansatz equation (\ref{BAE}) traces back to the 
eigenvalue formula of transfer matrix and the Bethe ansatz equation
 of Perk-Schultz model[Sc] except the vacuum part,
 some gauge factors and extra signs after some redefinition. 
 (See also, [Kul]). 

Let $\lambda \subset \mu$ be a skew-Young superdiagram labeled by 
the sequences of non-negative integers 
$\lambda =(\lambda_{1},\lambda_{2},\dots)$ and 
$\mu =(\mu_{1},\mu_{2},\dots)$ such that
$\mu_{i} \ge \lambda_{i}: i=1,2,\dots;$  
$\lambda_{1} \ge \lambda_{2} \ge \dots \ge 0$;  
$\mu_{1} \ge \mu_{2} \ge \dots \ge 0$ and 
$\lambda^{\prime}=(\lambda_{1}^{\prime},\lambda_{2}^{\prime},\dots)$ 
be the conjugate of $\lambda $ 
(see, figure\ref{young} and \ref{conjyoung}).
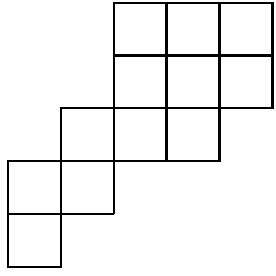
\begin{figure}
  \begin{center}
    \setlength{\unitlength}{2pt}
    \begin{picture}(50,55) 
      \put(0,0){\line(0,1){20}}
      \put(10,0){\line(0,1){30}}
      \put(20,10){\line(0,1){40}}
      \put(30,20){\line(0,1){30}}
      \put(40,20){\line(0,1){30}}
      \put(50,30){\line(0,1){20}}
      \put(0,0){\line(1,0){10}}      
      \put(0,10){\line(1,0){20}}
      \put(0,20){\line(1,0){40}}
      \put(10,30){\line(1,0){40}}
      \put(20,40){\line(1,0){30}}
      \put(20,50){\line(1,0){30}}
    \end{picture}
  \end{center}
  \caption{Young superdiagram with shape $\lambda \subset \mu$ : 
  $\lambda=(2,2,1,0,0)$, $\mu=(5,5,4,2,1)$}
  \label{young}
\end{figure}
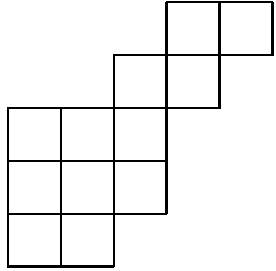
\begin{figure}
  \begin{center}
    \setlength{\unitlength}{2pt}
    \begin{picture}(50,55) 
      \put(0,0){\line(0,1){30}}
      \put(10,0){\line(0,1){30}}
      \put(20,0){\line(0,1){40}}
      \put(30,10){\line(0,1){40}}
      \put(40,30){\line(0,1){20}}
      \put(50,40){\line(0,1){10}}
      \put(0,0){\line(1,0){20}}      
      \put(0,10){\line(1,0){30}}
      \put(0,20){\line(1,0){30}}
      \put(0,30){\line(1,0){40}}
      \put(20,40){\line(1,0){30}}
      \put(30,50){\line(1,0){20}}
    \end{picture}
  \end{center}
  \caption{Young superdiagram with shape $\lambda ^{\prime} \subset
   \mu^{\prime}$ : $\lambda^{\prime}=(3,2,0,0,0)$,
    $\mu^{\prime}=(5,4,3,3,2)$}
    \label{conjyoung}
\end{figure}
On this skew-Young  superdiagram 
$\lambda \subset \mu$, we assign a coordinates $(i,j)\in {\bf Z}^{2}$ 
such that the row index $i$ increases as we go downwards and the column 
index $j$ increases as we go from left to right and that 
$(1,1)$ is on the top left corner of $\mu$.
Define an admissible tableau $b$ 
on the skew-Young superdiagram 
$\lambda \subset \mu$ as a set of element $b(i,j)\in J$ 
 labeled by the coordinates 
$(i,j)$ mentioned above, obeying the following rule 
(admissibility conditions).
\begin{enumerate}
\item 
For any elements of $J_{+}$,
\begin{numparts} 
\begin{equation}
  b(i,j) \prec b(i+1,j)
\end{equation}
\item 
 for any elements of $J_{-}$,
\begin{equation}
 b(i,j) \prec b(i,j+1),
\end{equation}
\item 
and for any elements of $J$,
\begin{equation}   
b(i,j) \preceq b(i,j+1),\quad b(i,j) \preceq b(i+1,j).
\end{equation}
\end{numparts}
\end{enumerate}
Let $B(\lambda \subset \mu)$ be 
the set of admissible tableaux 
 on $\lambda \subset \mu$.
For any skew-Young superdiagram $\lambda \subset \mu$, 
define the function ${\cal T}_{\lambda \subset \mu}(u)$ as follows
\begin{equation}
\fl {\cal T}_{\lambda \subset \mu}(u)=
 \frac{1}{{\cal F}_{\lambda \subset \mu}(u)} 
\sum_{b \in B(\lambda \subset \mu)}
\prod_{(i,j) \in (\lambda \subset \mu)}
(-1)^{p(b(i,j))}
z(b(i,j);u-\mu_{1}+\mu_{1}^{\prime}-2i+2j)	
\label{Tge1}
\end{equation}
where the product is taken over the coordinates $(i,j)$ on
 $\lambda \subset \mu$ and 
\begin{equation}
\fl {\cal F}_{\lambda \subset \mu}(u)= 
 \prod_{j=1}^{\mu_{1}}
{\cal F}^{\mu_{j}^{\prime}-\lambda_{j}^{\prime}}
(u+\mu_{1}^{\prime}-\mu_{1}-\mu_{j}^{\prime}-\lambda_{j}^{\prime}+2j-1).
\end{equation} 
In particular, for an empty diagram 
$\phi$, set ${\cal T}_{\phi}(u)={\cal F}_{\phi}(u)=1$. 
The following relations should be
 valid by the same reason mentioned in [KOS], 
that is, they will be verified by induction on $\mu_{1}$ 
or $\mu_{1}^{\prime}$.
\numparts
\begin{eqnarray}
\fl {\cal T}_{\lambda \subset \mu}(u)=det_{1 \le i,j \le \mu_{1}}
    ({\cal T}^{\mu_{i}^{\prime}-\lambda_{j}^{\prime}-i+j}
(u-\mu_{1}+\mu_{1}^{\prime}-\mu_{i}^{\prime}-\lambda_{j}^{\prime}+i+j-1))	
	\label{Jacobi-Trudi1} \\ 
	\fl =det_{1 \le i,j \le \mu_{1}^{\prime}}
    ({\cal T}_{\mu_{j}-\lambda_{i}+i-j}
    (u-\mu_{1}+\mu_{1}^{\prime}+\mu_{j}+\lambda_{i}-i-j+1))	
	\label{Jacobi-Trudi2} 
\end{eqnarray}
\endnumparts
For example, $\lambda=\phi, \mu=(2^2), r=1,s=0$ case, we have  
\begin{eqnarray}
{\cal T}_{(2^2)}(u)=
\frac{1}{{\cal F}_{(2^2)}(u)}
  \left( \begin{array}{|c|c|}\hline 
     	1 & 1 \\ \hline
     	2 & 2 \\ \hline 
     \end{array}
   -\begin{array}{|c|c|}\hline 
     	1 & 1 \\ \hline
     	2 & 3 \\ \hline 
     \end{array}
  -
    \begin{array}{|c|c|}\hline 
     	1 & 2 \\ \hline
     	2 & 3 \\ \hline 
     \end{array}
   +\begin{array}{|c|c|}\hline 
     	1 & 3 \\ \hline
     	2 & 3 \\ \hline 
     \end{array} \right) \nonumber \\
    \fl =
P_1(u+2)P_1(u+4)\frac{Q_2(u-2)}{Q_2(u+2)}- 
P_1(u+2)P_1(u+4)
\frac{Q_1(u+1)Q_2(u-2)}{Q_1(u+3)Q_2(u+2)} \nonumber \\ 
- P_1(u+2)^2\frac{Q_1(u+5)Q_2(u-2)}{Q_1(u+3)Q_2(u+4)}+
P_1(u+2)^2\frac{Q_2(u-2)}{Q_2(u+4)} \\ 
=\begin{array}{|cc|} 
     	{\cal T}^{2}(u-1) & {\cal T}^{3}(u) \\
     	{\cal T}^{1}(u) & {\cal T}^{2}(u+1) \\  
     \end{array}
     \nonumber 
 \end{eqnarray}
 where 
\begin{eqnarray}
{\cal T}^{1}(u)
=\begin{array}{|c|}\hline 
     	1  \\ \hline 
     \end{array}
+\begin{array}{|c|}\hline 
     	2  \\ \hline 
     \end{array}
-\begin{array}{|c|}\hline 
     	3 \\ \hline 
     \end{array} \\ 
=P_1(u+2)\frac{Q_1(u-1)}{Q_1(u+1)}
+P_1(u)\frac{Q_1(u+3)Q_2(u)}{Q_1(u+1)Q_2(u+2)}
-P_1(u)\frac{Q_2(u)}{Q_2(u+2)}, \nonumber \\ 
{\cal T}^{2}(u)
=\frac{1}{{\cal F}^{2}(u)} \left(
    \begin{array}{|c|}\hline 
     	1  \\ \hline 
     	2 \\ \hline
     \end{array}
-   \begin{array}{|c|}\hline 
     	1  \\ \hline 
     	3 \\ \hline
     \end{array}
-    \begin{array}{|c|}\hline 
     	2  \\ \hline 
     	3 \\ \hline
     \end{array}
+    \begin{array}{|c|}\hline 
     	3  \\ \hline 
     	3 \\ \hline
     \end{array} \right) \\
=P_1(u+3)\frac{Q_2(u-1)}{Q_2(u+1)}
-P_1(u+3)\frac{Q_1(u)Q_2(u-1)}{Q_1(u+2)Q_2(u+1)} \nonumber \\ 
-P_1(u+1)\frac{Q_1(u+4)Q_2(u-1)}{Q_1(u+2)Q_2(u+3)}
+P_1(u+1)\frac{Q_2(u-1)}{Q_2(u+3)}, \nonumber \\
{\cal T}^{3}(u)
=\frac{1}{{\cal F}^{3}(u)} \left(
 -   \begin{array}{|c|}\hline 
     	1  \\ \hline 
     	2 \\ \hline
     	3 \\ \hline
     \end{array}
+   \begin{array}{|c|}\hline 
     	1  \\ \hline 
     	3 \\ \hline
     	3 \\ \hline
     \end{array}
+    \begin{array}{|c|}\hline 
     	2  \\ \hline 
     	3 \\ \hline
     	3 \\ \hline
     \end{array}
-    \begin{array}{|c|}\hline 
     	3  \\ \hline 
     	3 \\ \hline
     	3 \\ \hline
     \end{array}
\right) \\ 
=-P_1(u+4)\frac{Q_2(u-2)}{Q_2(u+2)}
+P_1(u+4)\frac{Q_1(u+1)Q_2(u-2)}{Q_1(u+3)Q_2(u+2)} \nonumber \\ 
+P_1(u+2)\frac{Q_1(u+5)Q_2(u-2)}{Q_1(u+3)Q_2(u+4)}
-P_1(u+2)\frac{Q_2(u-2)}{Q_2(u+4)}. \nonumber 
 \end{eqnarray}
Remark1: If we drop the $u$ dependence of (\ref{Jacobi-Trudi1}) and 
(\ref{Jacobi-Trudi2}), they reduce to classical 
Jacobi-Trudi and Giambelli formulae for $sl(r+1|s+1)$ [BB1,PT], 
which bring us classical (super) characters. \\ 
Remark2: In the case $\lambda =\phi$ and $s=-1$, $(\ref{Jacobi-Trudi1})$
and $(\ref{Jacobi-Trudi2})$  
correspond to quantum analogue of Jacobi-Trudi and Giambelli 
formulae for $sl_{r+1}$ [BR].\\
Remark3:  $(\ref{Jacobi-Trudi1})$ and $(\ref{Jacobi-Trudi2})$ 
have the same form as 
the quantum Jacobi-Trudi and Giambelli formulae for 
$U_{q}(B_{n}^{(1)})$ in [KOS], but the function 
${\cal T}^{a}(u)$ is quite different. 

The following Theorem is essential in analytic Bethe ansatz, 
which can be proved along the 
similar line of the proof of Theorem 3.3.1. in [KS1].
\begin{theorem}\label{polefree}
For any integer $a$, the function ${\cal T}^a(u)$  
is free of poles under the condition that
the Bethe ansatz equation (\ref{BAE}) is valid.   
\end{theorem}
At first, we present a lemma which is necessary for the proof of 
the Theorem\ref{polefree}. Lemma\ref{lemma2box} is $sl(r+1|s+1)$ version 
of Lemma3.3.2.in [KS1] and follows straightforwardly from 
the definitions of $z(a;u)$ (\ref{z+}).
\begin{lemma} \label{lemma2box}
For any $b \in J_{+}-\{r+1\}$, the function
\begin{equation}
\begin{array}{|c|l}\cline{1-1}
    b   & _u \\ \cline{1-1} 
    b+1 & _{u-2}\\ \cline{1-1}
\end{array} \label{2box}
\end{equation}
does not contain the function $Q_b$ (\ref{Q_a}).
\end{lemma}
Proof of Theorem\ref{polefree}. 
For simplicity, we assume that the vacuum parts are formally trivial, 
that is, the left hand side of the 
 Bethe ansatz equation (\ref{BAE}) is constantly $-1$. 
We prove that ${\cal T}^a(u)$ is free of color $b$ pole, namely, 
$Res_{u=u_{k}^{(b)}+\cdots}{\cal T}^a(u)=0$ for any 
 $b \in J-\{r+s+2\}$
 under the condition that the Bethe
  ansatz equation (\ref{BAE}) is valid. 
 The function $z(c;u)=\framebox{$c$}_{u}$ with $c\in J $ has 
the color $b$ pole only for $c=b$ or $b+1$, so we shall trace only 
\framebox{$b$} or \framebox{$b+1$}.
Denote $S_{k}$ the partial sum of ${\cal T}^a(u)$, which contains 
k boxes among \framebox{$b$} or \framebox{$b+1$}.
 Apparently, $S_{0}$ does not have 
color $b$ pole. This is also the case with $S_{2}$ 
for $b\in J_{+}-\{r+1\}$ since the admissible tableaux have  
 the same subdiagrams as in (\ref{2box}) and thus do not involve $Q_{b}$
 by lemma\ref{lemma2box}. Now we examine $S_{1}$ which is the summation 
 of the tableaux of the form 
\begin{equation} 
\begin{array}{|c|}\hline
    \xi \\ \hline 
    b   \\ \hline 
   \zeta \\ \hline
\end{array}
\qquad \qquad 
\begin{array}{|c|}\hline
    \xi \\ \hline 
    b+1 \\ \hline 
   \zeta \\ \hline
\end{array} \label{tableaux1}
\end{equation}
where \framebox{$\xi$} and \framebox{$\zeta$} are columns with 
total length $a-1$ and they do not involve \framebox{$b$} and 
\framebox{$b+1$}. Thanks to the relations (\ref{res1}-\ref{res3}), 
color $b$ residues in these tableaux (\ref{tableaux1}) cancel each 
other under the Bethe ansatz equation (\ref{BAE}). Then we deal with 
$S_{k}$ only for $3 \le k \le a $ and $k=2$ with 
$b\in J_{-} \cup \{r+1 \}-\{r+s+2\}$ 
from now on. In this case, only the case for 
$b \in \{r+1 \} \bigcup J_{-}-\{r+s+2\}$
 should be considered because, in the case for 
 $b \in J_{+}- \{r+1 \}$, \framebox{$b$} or \framebox{$b+1$} appear at 
  most twice in one column.\\
The case $ b=r+1 $ : $S_{k} (k\ge 2)$ is the 
summation of the tableaux of the form  
\begin{equation}
\begin{array}{|c|l}\cline{1-1}
    \xi & \\ \cline{1-1} 
    r+1   & _v  \\ \cline{1-1} 
    r+2 & _{v-2} \\ \cline{1-1} 
   \vdots & \\  \cline{1-1} 
    r+2 & _{v-2k+2} \\ \cline{1-1} 
   \zeta & \\ \cline{1-1}
\end{array}
= \frac{Q_{r+1}(v+r-2k+1)Q_{r+2}(v+r)}{Q_{r+1}(v+r+1)Q_{r+2}(v+r+2)}X_{3} 
\label{tableauxk1}
\end{equation}
and 
\begin{equation}
\begin{array}{|c|l}\cline{1-1}
    \xi & \\ \cline{1-1} 
    r+2   & _v \\ \cline{1-1} 
    r+2   & _{v-2} \\ \cline{1-1} 
   \vdots & \\ \cline{1-1} 
    r+2 & _{v-2k+2}\\ \cline{1-1} 
   \zeta & \\ \cline{1-1}
\end{array}
=\frac{Q_{r+1}(v+r-2k+1)Q_{r}(v+r)}{Q_{r+1}(v+r+1)Q_{r}(v+r+2)} X_{3}
\label{tableauxk2}
\end{equation}
where \framebox{$\xi$} and \framebox{$\zeta$} are columns with 
total length $a-k$, which do not contain \framebox{$r+1$} and 
\framebox{$r+2$}; $v=u+h_1$: $h_1$ is some shift parameter; 
the function $X_{3}$ does not contain the function $Q_{r+1}$.
Obviously, color $b=r+1$ residues in the 
(\ref{tableauxk1}) and (\ref{tableauxk2}) cancel each other
 under the Bethe ansatz equation (\ref{BAE}). \\ 
The case $b \in J_{-}-\{r+s+2\}$:
 $S_{k} (k \ge 2)$ is the summation of 
 the tableaux of the form 
\begin{eqnarray} 
f(k,n,\xi,\zeta,u):=
\begin{array}{|c|l}\cline{1-1}
    \xi & \\ \cline{1-1} 
    b   & _v \\ \cline{1-1} 
    \vdots & \\ \cline{1-1} 
    b & _{v-2n+2}\\ \cline{1-1} 
    b+1 & _{v-2n} \\ \cline{1-1} 
   \vdots & \\ \cline{1-1} 
    b+1 & _{v-2k+2}\\ \cline{1-1} 
   \zeta & \\ \cline{1-1}
\end{array}  \nonumber \\
=\frac{Q_{b-1}(v+2r+3-2n-b)Q_{b}(v+2r+4-b)}
      {Q_{b-1}(v+2r+3-b)Q_{b}(v+2r+4-2n-b)} \\ 
  \times \frac{Q_{b}(v+2r+2-2k-b)Q_{b+1}(v+2r+3-2n-b)}
          {Q_{b}(v+2r+2-2n-b)Q_{b+1}(v+2r+3-2k-b)} X_{4} 
,\quad  0 \le n \le k \nonumber
\label{tableauxk3}
\end{eqnarray}
where \framebox{$\xi$} and \framebox{$\zeta$} are columns with 
total length $a-k$, which do not contain \framebox{$b$} and 
\framebox{$b+1$}; $v=u+h_2$: $h_2$ is some shift parameter and 
is independent of $n$; the function $X_{4}$ does not have color $b$
 pole and is independent of $n$.
$f(k,n,\xi,\zeta,u)$ has color $b$ poles at
 $u=-h_2-2r-2+b+2n+u_{p}^{(b)}$ and $u=-h_2-2r-4+b+2n+u_{p}^{(b)}$
  for $1 \le n \le k-1$; at $u=-h_2-2r-2+b+u_{p}^{(b)}$ 
for $n=0$ ; at $u=-h_2-2r-4+b+2k+u_{p}^{(b)}$ for $n=k$. 
Evidently, color $b$ residue at 
$u=-h_2-2r-2+b+2n+u_{p}^{(b)}$
 in  $f(k,n,\xi,\zeta,u)$ and $f(k,n+1,\xi,\zeta,u)$
 cancel each other under the Bethe ansatz equation (\ref{BAE}). 
 Thus, under the Bethe ansatz equation
  (\ref{BAE}), $\sum_{n=0}^{k}f(k,n,\xi,\zeta,u)$ 
 is free of color $b$ poles, so is $S_{k}$.
\rule{5pt}{10pt} \\ 
Applying Theorem\ref{polefree} to  (\ref{Jacobi-Trudi1}),
 one can show that 
${\cal T}_{\lambda \subset \mu}(u)$ 
is free of poles under the Bethe ansatz equation (\ref{BAE}). 
The function ${\cal T}_{\lambda \subset \mu}(u)$  
 should express 
the eigenvalue of the transfer matrix whose auxiliary space 
$W_{\lambda \subset \mu}(u)$ is labeled by the skew-Young superdiagram 
 with shape 
$\lambda \subset \mu$.   We assume that $W_{\lambda \subset \mu}(u)$ 
is a finite dimensional module of the super Yangian $Y(sl(r+1|s+1))$ [N] 
( or quantum super affine algebra $U_{q}(sl(r+1|s+1)^{(1)})$ [Y] in 
 the trigonometric case ). 
 On the other hand, for  $\lambda =\phi $  case, highest  
 weight representation of Lie superalgebra $sl(r+1|s+1)$, 
 which is a classical counterpart of $W_{\mu}(u)$, is
  characterized by the highest weight 
 whose Kac-Dynkin labels $a_{1},a_{2},\dots ,a_{r+s+1}$ [BMR] are 
 given as follows:
\begin{eqnarray}
 a_{j}=\mu_{j}-\mu_{j+1} \quad {\rm for} \quad  
  1 \le j \le r \nonumber\\
 a_{r+1}=\mu_{r+1}+\eta_{1}  \label{KacDynkin}    \\
 a_{j+r+1}=\eta_{j}-\eta_{j+1} \quad {\rm for} \quad   
 1 \le j \le s \nonumber
\end{eqnarray}
where $\eta_{j}=max\{\mu_{j}^{\prime}-r-1,0 \}$; $\mu_{r+2} \le s+1$ 
for covariant case. One can read the relations (\ref{KacDynkin})  
 from the \symbol{96}top term\symbol{39} [KS1,KOS] in (\ref{Tge1})
  for large $q^u$ 
 (see, figure\ref{top}). The \symbol{96}top term\symbol{39} in
  (\ref{Tge1}) is the term labeled by the tableau $b$ such that 
\begin{equation}
b(i,j)=
 \left\{
         \begin{array}{@{\,}ll}
             i  &    \mbox{ for } \quad 1 \le j \le \mu_{i} \quad  
                     \mbox{ and } \quad 1 \le i \le r+1 \\ 
             r+j+1 & \mbox{ for } \quad 1 \le j \le \mu_{i} \quad
                     \mbox{ and } \quad r+2 \le i \le \mu_{1}^{\prime} . 
          \end{array}
 \right. 	
\end{equation}
\begin{figure}
  \begin{center}
    \setlength{\unitlength}{2pt}
    \begin{picture}(50,60) 
      \put(0,0){\line(0,1){60}}
      \put(10,0){\line(0,1){60}}
      \put(20,10){\line(0,1){50}}
      \put(30,30){\line(0,1){30}}
      \put(40,40){\line(0,1){20}}
      \put(50,50){\line(0,1){10}}
      \put(0,0){\line(1,0){10}}      
      \put(0,10){\line(1,0){20}}
      \put(0,20){\line(1,0){20}}
      \put(0,30){\line(1,0){30}}
      \put(0,40){\line(1,0){40}}
      \put(0,50){\line(1,0){50}}
      \put(0,60){\line(1,0){50}}
      \put(4,4){4}
      \put(4,14){4}
      \put(4,24){4}
      \put(4,34){3}
      \put(4,44){2}
      \put(4,54){1}
      \put(14,14){5}
      \put(14,24){5}
      \put(14,34){3}
      \put(14,44){2}
      \put(14,54){1}
      \put(24,34){3}
      \put(24,44){2}
      \put(24,54){1}
      \put(34,44){2}
      \put(34,54){1}
      \put(44,54){1}
    \end{picture}
  \end{center}
  \caption{Young supertableau corresponding to the top term 
  for $sl(3|2)$; $\lambda \subset \mu$ : 
  $\lambda=\phi$, $\mu=(5,4,3,2,2,1)$}
  \label{top}
\end{figure}
 Then, for large $q^u$, we have   
\begin{eqnarray}
  \prod_{(i,j) \in \mu}
  (-1)^{p(b(i,j))}
  z(b(i,j);u+\mu_{1}^{\prime}-\mu_{1}-2i+2j) \nonumber \\  
 = (-1)^{\sum_{i=r+2}^{\mu_{1}^{\prime}}\mu_{i}} 
\left\{ \prod_{i=1}^{r+1} \prod_{j=1}^{\mu_{i}}
  z(i;u+\mu_{1}^{\prime}-\mu_{1}-2i+2j) \right\} \nonumber \\ 
 \times 
 \left\{ \prod_{j=1}^{\mu_{r+2}} \prod_{i=r+2}^{\mu_{j}^{\prime}}
  z(r+j+1;u+\mu_{1}^{\prime}-\mu_{1}-2i+2j) \right\} \nonumber \\  
 \approx (-1)^{\sum_{i=r+2}^{\mu_{1}^{\prime}}\mu_{i}}  
 q^{-2\sum N_{b}a_{b}t_{b}}.
\end{eqnarray}
Here we omit the vacuum part $\psi_{a}$. 
The \symbol{96}top term\symbol{39} is considered to be related 
with the \symbol{96}highest weight vector\symbol{39}. 
See [KS1,KOS], for more details.
\section{Functional equations}
Consider the following Jacobi identity: 
\begin{equation}
\fl  {D}\left[
   \begin{array}{c}
        b \\
        b 
   \end{array}
  \right]
  {D}
   \left[
   \begin{array}{c}
        c \\ 
        c 
   \end{array}
  \right]-
 {D}\left[
   \begin{array}{c}
        b \\
        c 
   \end{array}
  \right]
 {D}\left[
   \begin{array}{c}
        c \\
        b 
   \end{array}
  \right]=
 {D}\left[
   \begin{array}{cc}
        b & c\\
        b & c
   \end{array}
  \right]
   {D},  
   \quad b \ne c    
        \label{jacobi}	
\end{equation} 
where $D$ is the determinant of a matrix and 
${D}\left[
   \begin{array}{ccc}
        a_{1} & a_{2} & \dots \\
        b_{1} & b_{2} & \dots
   \end{array}
   \right]$
is its minor removing $a_{\alpha}$'s rows and 
$b_{\beta}$'s columns.
Set $\lambda = \phi$, $ \mu =(m^a)$ in (\ref{Jacobi-Trudi1}).
 From the relation (\ref{jacobi}), we have
\begin{equation}
\fl  {\cal T}_{m}^{a}(u-1) {\cal T}_{m}^{a}(u+1)  = 
    {\cal T}_{m+1}^{a}(u) {\cal T}_{m-1}^{a}(u)+
    g_{m}^{a}(u) {\cal T}_{m}^{a-1}(u) {\cal T}_{m}^{a+1}(u) 
        \label{t-sys1} 
\end{equation}
where $a,m \ge 1$; ${\cal T}_{m}^{a}(u)={\cal T}_{(m^a)}(u)$: $a,m \ge 1$;
 ${\cal T}_{m}^{0}(u)=1$: $m \ge 0$; ${\cal T}_{0}^{a}(u)=1$: $a \ge 0$; 
$g_{m}^{1}(u)=\prod_{j=1}^{m} P_{1}(u-m+2j-2)$:$m \ge 1$; 
$g_{m}^{a}(u)=1$: $a \ge 2$ and $m \ge 0$, or $a=1$ and $m=0$. 
Note that the following relation holds: 
\begin{equation}
 g_{m}^{a}(u+1)g_{m}^{a}(u-1)=g_{m+1}^{a}(u)g_{m-1}^{a}(u) \quad {\rm for }
 \quad a,m \ge 1.
\end{equation} 
The functional equation (\ref{t-sys1}) is a special 
case of Hirota bilinear difference equation [H]. 
In addition, there are some restrictions on it,  
which we consider below.

\begin{theorem}\label{vanish} 
 ${\cal T}_{\lambda \subset \mu}(u)=0$ 
if $\lambda \subset \mu$ contains 
a rectangular subdiagram with $r+2$ rows and $s+2$ columns.
(see, [DM,MR])
\end{theorem}
Proof. We assume the coordinate of the top left corner of this subdiagram 
 is $(i_1,j_1)$. Consider the tableau $b$ on this Young superdiagram 
 $\lambda \subset \mu$.
Fill the first column of this subdiagram from the top to the bottom by the 
elements of $b(i,j_1) \in J$: $i_1 \le i \le i_1+r+1$, 
so as to meet the admissibility conditions 
(i), (ii) and (iii).
We find $b(i_1+r+1,j_1) \in J_{-}$. Then we have  
 $r+2 \preceq b(i_1+r+1,j_1) \prec b(i_1+r+1,j_1+1)
  \prec \dots \prec b(i_1+r+1,j_1+s+1) $. 
 This contradicts the condition $b(i_1+r+1,j_1+s+1) \preceq r+s+2$.
  \rule{5pt}{10pt} \\
As a corollary, we have 
\begin{equation}
{\cal T}_{m}^{a}(u)=0 \quad {\rm for} \quad 
a \ge r+2  \quad {\rm and} \quad  m \ge s+2. \label{vanish2}
\end{equation}  
Consider the admissible tableaux on the Young superdiagram with shape 
$(m^{r+1})$. From the admissibility conditions 
(i), (ii) and (iii), only such tableaux as 
 $b(i,j)=i$ for $1 \le i \le r+1$ and $1\le j \le m-s-1$ are admissible. 
Then we have, 
\begin{numparts}
\begin{eqnarray}
\fl {\cal T}_{m}^{r+1}(u)={\cal T}_{(m^{r+1})}(u) \nonumber \\ 
\fl= \frac{1}{{\cal F}_{(m^{r+1})}(u)} \sum_{b \in B(m^{r+1})}
\prod_{(i,j) \in (m^{r+1})}
(-1)^{p(b(i,j))}
z(b(i,j);u+r+1-m-2i+2j)	\nonumber \\ 
\fl=\frac{1}{{\cal F}_{(m^{r+1})}(u)} \prod_{i=1}^{r+1}
\prod_{j=1}^{m-s-1}
(-1)^{p(i)}  z(i;u+r+1-m-2i+2j) \nonumber \\
\fl \times
\sum_{b \in B((s+1)^{r+1})} 
\prod_{i=1}^{r+1}
\prod_{j=m-s}^{m}
(-1)^{p(b(i,j))}
z(b(i,j);u+r+1-m-2i+2j) \nonumber \\ 
\fl={\cal F}^{m-s}(u+r-s+2)
\frac{Q_{r+1}(u-m)}{Q_{r+1}(u+m-2s-2)}
\times
{\cal T}_{s+1}^{r+1}(u+m-s-1),\label{red1} \\
 \quad m \ge s+1.  \nonumber
\end{eqnarray}
Similarly, we have
\begin{eqnarray}
\fl	{\cal T}_{s+1}^{a}(u)  =
      (-1)^{(s+1)(a-r-1)}
	  \frac{Q_{r+1}(u-a-s+r)}{Q_{r+1}(u+a-s-r-2)} 
	  \times 
	  {\cal T}_{s+1}^{r+1}(u+a-r-1) ,
	  \nonumber \\ \quad a \ge r+1. 
	\label{red2} 
\end{eqnarray}
\end{numparts}
From the relations (\ref{red1}) and (\ref{red2}), we obtain
\begin{theorem}\label{dual} For $a \ge 1$ and $r \ge 0$,
 the following relation is valid. 
\begin{equation} 
{\cal T}_{a+s}^{r+1}(u)=(-1)^{(s+1)(a-1)}
{\cal F}^{a}(u+r-s+2) {\cal T}_{s+1}^{r+a}(u).
\end{equation}
\end{theorem}
Applying the relation (\ref{vanish2}) to (\ref{t-sys1}),
 we obtain 
\begin{numparts}
\begin{equation}
\fl  {\cal T}_{m}^{r+1}(u-1) {\cal T}_{m}^{r+1}(u+1)  = 
    {\cal T}_{m+1}^{r+1}(u) {\cal T}_{m-1}^{r+1}(u)
     \quad m \ge s+2,
        \label{laplace1} 
\end{equation}
\begin{equation}
\fl  {\cal T}_{s+1}^{a}(u-1) {\cal T}_{s+1}^{a}(u+1)  = 
    g_{s+1}^{a}(u) {\cal T}_{s+1}^{a-1}(u) {\cal T}_{s+1}^{a+1}(u)
     \quad a \ge r+2.
        \label{laplace2} 
\end{equation}
\end{numparts}
Thanks to Theorem\ref{dual}, (\ref{laplace1}) is equivalent 
to (\ref{laplace2}).
From Theorem \ref{dual} , we also have
\begin{eqnarray}
 \fl {\cal T}_{s+1}^{r+1}(u-1) {\cal T}_{s+1}^{r+1}(u+1)  = 
    {\cal T}_{s+2}^{r+1}(u)({\cal T}_{s}^{r+1}(u)+
    (-1)^{s+1}\frac{{\cal T}_{s+1}^{r}(u)}{{\cal F}^{2}(u+r-s+2)}).
\end{eqnarray}
Remark: In the relation (\ref{red1}), we assume that the parameter 
$m$ takes only integer value. 
However, there is a possibility of $m$ taking non-integer values, 
 except some \symbol{96}singular point\symbol{39},
  for example, on which 
  right hand side of (\ref{red1}) contains constant terms, 
by \symbol{96}analytic continuation\symbol{39}. 
We can easily observe this fact from the right hand side of (\ref{red1}) 
as long as normalization factor ${\cal F}^{m-s}(u)$ is disregarded. 
This seems to correspond to the fact that r+1 th Kac-Dynkin label 
(\ref{KacDynkin}) $a_{r+1}$ can take non-integer value [Ka].
Furthermore, these circumstances seem to be connected with the lattice 
models based upon the  
solution of the graded Young-Baxter equation, which depends on 
 non-additive continuous parameter (see for example, [M,PF]). 
\section{Summary and discussion}
In this paper, we have executed analytic Bethe ansatz for Lie superalgebra
$sl(r+1|s+1)$. Pole-freeness of eigenvalue formula of transfer matrix in 
 dressed vacuum form was shown for a 
wide class of finite dimensional representations labeled by
 skew-Young superdiagrams.
 Functional relation has been given especially for the eigenvalue formulae 
 of transfer matrices 
 in dressed vacuum form 
 labeled by rectangular Young superdiagrams, 
 which is a special case of Hirota bilinear difference equation with  
 some restrictive relations.

It should be emphasized that our method presented in this paper is also 
applicable  
even if such factors like extra sign (different from that of(\ref{BAE})), 
gauge factor, etc. appear in the Bethe ansatz equation (\ref{BAE}). 
This is because such factors do not affect the analytical
 property of right 
 hand side of the Bethe ansatz equation (\ref{BAE}). 

It would be an interesting problem to extend similar analyses to 
 mixed representation cases [BB2].
  So far we have only found several determinant 
 representations of mixed tableau. The simplest one is given as follows.
 \begin{equation}
 \sum_{(a,b)\in X} (-1)^{p(a)+p(b)}\dot{z}(a;u+s)z(b;u+r)=
     \begin{array}{|cc|}
        \dot{{\cal T}}^{1}(u+s) & 1        \\
         1          & {\cal T}^{1}(u+r)
     \end{array}
     \label{mix}
 \end{equation}
 where $X=\{ (a,b): a\in \dot{J}; b \in J;(a,b) \ne (-1,1) \}$ 
 for $sl(r+1|s+1): r \ne s$; 
  $\dot{{\cal T}}^{1}(u)$ and $\dot{J}$ are the expressions 
 related to contravariant representations (see, Appendix B). 
 Here we assume that the vacuum parts are formally trivial. 
Note that (\ref{mix}) reduces to the classical one for 
$sl(r+1|s+1); r \ne s$ [BB2], if we drop the $u$ dependence.
 
In this paper, we mainly consider  the Bethe ansatz equations 
for distinguished root system. 
The case for non-distinguished root system will be  
achieved by some modifications of the set $J_{+}$, $J_{-}$
and the function $z(a;u)$ without changing 
 the set $J$ and tableau sum rule (see, Appendix C,D).  
It will be interesting to extend a similar analysis  
presented in this paper for other Lie superalgebras, such 
as $osp(m|2n)$.
\ack 
The author would like to thank Professor A Kuniba for continual 
encouragement, useful advice and comments on the manuscript.
He also thanks Dr J Suzuki for helpful discussions and 
 pointing out some mistake in the earlier version of the manuscript;     
 Professor T Deguchi for useful comments. 
\appendix
\section{Example of the $L$ operator and transfer matrix}
In this section, we define the transfer matrix along the same line 
presented in [EK]. 
Let $L(u)_{\alpha \beta}^{a b}$ be the $L$ operator [KulSk,PS1,PS2,Sc,BS] 
such that 
\begin{equation}
\fl L(u)_{a a}^{a a}=[u+2(-1)^{p(a)}], L(u)_{a a}^{b b}=[u], 
L(u)_{a b}^{b a}=[2(-1)^{p(a)p(b)}]q^{sign(a-b)u}
\end{equation}
where we assume $a \ne b;a,b \in J$. 
The monodromy matrix ${\cal J}(u)$ is defined as 
\begin{eqnarray}
\fl {\cal J}(u)_{b, \beta_1 \dots \beta_N}^{a, \gamma_1 \dots \gamma_N}
=\sum_{a_1,\dots ,a_N} L(u)_{\gamma_N \beta_N}^{a a_N} 
  L(u)_{\gamma_{N-1} \beta_{N-1}}^{a_N a_{N-1}} \cdots 
  L(u)_{\gamma_2 \beta_2}^{a_2 a_1}
  L(u)_{\gamma_1 \beta_1}^{a_1 b} \\ 
 \times (-1)^{\sum_{i=2}^{N}(p(\gamma_i)+p(\beta_i)) 
  \sum_{j=1}^{i-1} p(\gamma_j)} \nonumber 
\end{eqnarray}
The transfer matrix is defined as supertrace of the monodromy matrix 
\begin{equation}
t(u)_{\beta_1 \dots \beta_N}^{\gamma_1 \dots \gamma_N}=
\sum_{a=1}^{r+s+2} (-1)^{p(a)}
{\cal J}(u)_{a, \beta_1 \dots \beta_N}^{a, \gamma_1 \dots \gamma_N}
\end{equation}
Thanks to the intertwining relation, the commutativity 
relation $[t(u),t(v)]=0$ follows. 
The function ${\cal T}^{1}(u)$ defined in 
(\ref{generating}) will coincide with the 
the spectrum of the transfer matrix 
$t(u)$ under the Bethe ansatz equation (\ref{BAE}) 
for relevant $N_j$.
For example, for $r=0,s=1$, $N-N_1,N_1-N_2,N_2$ 
(see, (\ref{Q_a})) denote the 
number of $\gamma_i$ equal to 1, 2, 3 in the set 
$\{ \gamma_{1},\dots,\gamma_{N} \}$ respectively. 
Moreover, the function 
\begin{equation}
 {\cal T}^{1}(u)=
    \begin{array}{|c|}\hline 
     	1 \\ \hline 
     \end{array}
   -
    \begin{array}{|c|}\hline 
     	2 \\ \hline 
     \end{array}
   -\begin{array}{|c|}\hline 
     	3 \\ \hline 
     \end{array} 
\end{equation}
coincides with Sutherland's solution [Su] on supersymmetric $t-J$ model 
presented in [EK] 
in the limit $q \to 1$ 
except overall scalar factor after some redefinition.
\section{On the expressions related to contravariant representations}
In the main text, we have treated mainly the expressions 
related to covariant representations.  
For contravariant representations, we can also play a similar game. 
We often mark the expression related to 
contravariant representation with a dot.
In contravariant case, the relations (\ref{set}), (\ref{order}), 
(\ref{grading}), (\ref{z+}) and (\ref{KacDynkin})
become respectively as follows:
\begin{eqnarray}
\dot{J}=\{ -1,-2,\dots,-r-s-2\} , 
\dot{J}_{+}=\{ -1,-2,\dots,-r-1\},
\label{set2}  \\
\dot{J}_{-}=\{ -r-2,-r-3,\dots,-r-s-2\}, \nonumber
\end{eqnarray}
\begin{equation}
-r-s-2\prec -r-s-1 \prec \cdots \prec -1, 
\end{equation}
\begin{equation}
      p(a)=\left\{
              \begin{array}{@{\,}ll}
                0  & \mbox{for $a \in \dot{J}_{+}$} \\ 
                1 & \mbox{for $a \in \dot{J}_{-}$}
                 \quad 
              \end{array}
            \right. ,
\end{equation}
\begin{eqnarray}
\fl  \dot{z}(a;u)=\psi_{a}(u)
      \frac{Q_{-a-1}(u+r-s+a-1)Q_{-a}(u+r-s+a+2)}
       {Q_{-a-1}(u+r-s+a+1)Q_{-a}(u+r-s+a)} 
    \qquad {\rm for} \quad a \in \dot{J}_{+},
        \nonumber \\
\fl    \dot{z}(a;u)=\psi_{a}(u)
       \frac{Q_{-a-1}(u-r-s-a-1)Q_{-a}(u-r-s-a-4)}
        {Q_{-a-1}(u-r-s-a-3)Q_{-a}(u-r-s-a-2)} 
    \qquad {\rm for} \quad  a \in \dot{J}_{-}, \label{dotz}
\end{eqnarray}
\begin{eqnarray}
 a_{r+1-j}=\xi_{j}-\xi_{j+1} \quad {\rm for}
  \quad   1 \le j \le r, \nonumber  \\
 a_{r+1}=-\xi_{1}-\dot{\mu}_{s+1}^{\prime},                    \\
 a_{r+s+2-j}=\dot{\mu}_{j}^{\prime}-\dot{\mu}_{j+1}^{\prime} 
 \quad {\rm for} \quad   1 \le j \le s,  \nonumber \label{kacdynkcot}
\end{eqnarray}
where $\xi_{j}=max\{\dot{\mu}_{j}-s-1,0 \} $; 
$\dot{\mu}_{s+2}^{\prime} \le r+1$. 
The function (\ref{drinfeld}) and (\ref{psi}) take the form 
 \begin{equation}
\fl P_{a}^{(j)}(u)=[u-w_{j}]^{\delta_{a,r+s+1}}, \quad 
 \psi_{a}(u)=
   \left\{
    \begin{array}{@{\,}ll}
      P_{r+s+1}(u-2) & \mbox{for } \quad a=-r-s-2 \\ 
      P_{r+s+1}(u) & \mbox{for } \quad a \in \dot{J}-\{-r-s-2\}
    \end{array} 
   \right. 
 \end{equation}
 if the quantum space is labeled by the contravariant Young superdiagram
  with shape 
 $\dot{\tilde{\mu}}=(1^1)$; 
 \begin{equation}
\fl P_{a}^{(j)}(u)=[u-w_{j}]^{\delta_{a,1}}, \quad 
 \psi_{a}(u)=
   \left\{
    \begin{array}{@{\,}ll}
      P_{1}(u+r-s-2) & \mbox{for } \quad a=-1 \\ 
      P_{1}(u+r-s) & \mbox{for } \quad a \in \dot{J}-\{-1\}
    \end{array} \label{psi2}
   \right. 
 \end{equation}
 if the quantum space is labeled by the covariant Young superdiagram
  with shape $\tilde{\mu}=(1^1)$.
  
  If the quantum space is labeled by the contravariant Young superdiagram,
  in contrast to covariant case, 
  the parameter $t_{r+1}$ in left hand side of the 
  Bethe ansatz equation (\ref{BAE}) will be $-1$, since
  $r+1$ th Kac-Dynkin label takes negative 
  value for contravariant Young superdiagram [BMR].
 For $-a \in \dot{J}$ and (\ref{dotz}) with (\ref{psi2}),
  the following relation holds 
 \begin{equation}
  z(a;u)=(-1)^N\dot{z}(-a;s-r-u)|_{u_k^{(a)} \to -u_k^{(a)}, w_i \to -w_i}.
 \end{equation}
 Note that this relation reduces to the crossing symmetry [R2] for 
 $sl_{r+1}$, if we set $s=-1$ (see, also [KS1]).
  Pole freeness of the function  
$\dot{{\cal T}}_{\dot{\lambda} \subset \dot{\mu}}(u)$
 under the Bethe ansatz equation (\ref{BAE}) 
can be proved in the same way as {\it Theorem}\ref{polefree}.
\section{Example of non-distinguished simple roots case :
$p(1)=1,p(2)=0,p(3)=1$ grading}
\begin{figure}
    \setlength{\unitlength}{0.75pt}
    \begin{center}
    \begin{picture}(80,30) 
      \put(10,20){\circle{20}}
      \put(2.929,12.9289){\line(1,1){14.14214}}
      \put(2.929,27.07107){\line(1,-1){14.14214}}
      \put(20,20){\line(1,0){40}}
      \put(70,20){\circle{20}}
      \put(62.929,12.9289){\line(1,1){14.14214}}
      \put(62.929,27.07107){\line(1,-1){14.14214}}
      \put(6,0){$\alpha_{1}$}
      \put(66,0){$\alpha_{2}$}
  \end{picture}
  \end{center}
  \caption{Dynkin diagram for the Lie superalgebra
  $sl(1|2)$ corresponding to the non-distinguished simple roots :
   $ \deg(\alpha_1)= \deg(\alpha_2)=1$.}
  \label{dinkynno1}
\end{figure}
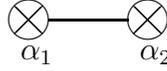

Let $\alpha_{1}$ and $\alpha_{2}$ be the simple roots of $sl(1|2)$ 
normalized so that 
 $(\alpha_{1}|\alpha_{1})=(\alpha_{2}|\alpha_{2})=0$ and 
 $(\alpha_{1}|\alpha_{2})=(\alpha_{2}|\alpha_{1})=-1$
  (see figure \ref{dinkynno1}).
In this case, the sets (\ref{set}) and (\ref{set2}) become 
$J_{+}=\{ 2\}$, $J_{-}=\{1, 3 \}$;$ \dot{J}_{+}=\{ -2\}$,  
$\dot{J}_{-}=\{-1, -3 \}$.
The function $z(a;u)=\framebox{$a$}_{u}$ $(a \in J)$ has the form  
\begin{equation}
\fl \framebox{1}=[u-2]^N \frac{Q_{1}(u+1)}{Q_{1}(u-1)}, 
\framebox{2}=[u]^N \frac{Q_{1}(u+1)Q_{2}(u-2)}{Q_{1}(u-1)Q_{2}(u)}, 
 \framebox{3}=[u]^N \frac{Q_{2}(u-2)}{Q_{2}(u)}  
\end{equation}
and the function  
$\dot{z}(a;u)=\framebox{$a$}_{u}$ $(a \in \dot{J})$ has the form  
\begin{equation}
\fl \framebox{-3}=[u-2]^N \frac{Q_{2}(u+1)}{Q_{2}(u-1)},
\framebox{-2}=[u]^N \frac{Q_{1}(u-2)Q_{2}(u+1)}{Q_{1}(u)Q_{2}(u-1)}, 
\framebox{-1}=[u]^N \frac{Q_{1}(u-2)}{Q_{1}(u)}.
\end{equation}
Here we assume the quantum spaces are labeled by Young superdiagrams 
with shapes 
$\tilde{\mu} =(1^1)$ and $\dot{\tilde{\mu}}=(1^1)$ respectively;
 for simplicity, 
 inhomogeneity parameters $w_i$ are set to $0$. 
For example, for $\lambda =\phi;\mu=(2^1)$, (\ref{Tge1}) has the form 
\begin{eqnarray}
  {\cal T}_{2}^{1}(u)=
   - \begin{array}{|c|c|}\hline 
     	1 & 2 \\ \hline 
     \end{array}
   +
    \begin{array}{|c|c|}\hline 
     	1 & 3 \\ \hline 
     \end{array}
   +\begin{array}{|c|c|}\hline 
     	2 & 2 \\ \hline 
     \end{array}
   -\begin{array}{|c|c|}\hline 
     	2 & 3 \\ \hline 
     \end{array} \nonumber \\ 
\fl =
-[u-3]^N [u+1]^N\frac{Q_{1}(u+2)Q_{2}(u-1)}{Q_{1}(u-2)Q_{2}(u+1)}+
[u-3]^N [u+1]^N\frac{Q_{1}(u)Q_{2}(u-1)}{Q_{1}(u-2)Q_{2}(u+1)}
 \nonumber \\ 
+[u-1]^N [u+1]^N\frac{Q_{1}(u+2)Q_{2}(u-3)}{Q_{1}(u-2)Q_{2}(u+1)} \\ 
 -[u-1]^N [u+1]^N\frac{Q_{1}(u)Q_{2}(u-3)}{Q_{1}(u-2)Q_{2}(u+1)}
 \nonumber  
\end{eqnarray}
and for $\lambda =\phi; \mu=(1^2)$, (\ref{Tge1}) has the form 
\begin{eqnarray}
  {\cal T}_{1}^{2}(u)=
  \frac{1}{[u-1]^N}
  \left(
     \begin{array}{|c|}\hline 
     	1 \\ \hline 
     	1 \\ \hline
     \end{array}
  - 
    \begin{array}{|c|}\hline 
     	1 \\ \hline 
     	2 \\ \hline
     \end{array}
  + 
  \begin{array}{|c|}\hline 
     	1 \\ \hline 
     	3 \\ \hline
     \end{array}
  - 
  \begin{array}{|c|}\hline 
     	2 \\ \hline 
     	3 \\ \hline
    \end{array}
  + 
  \begin{array}{|c|}\hline 
     	3 \\ \hline 
     	3 \\ \hline
     \end{array} \right) \nonumber \\ 
  =
[u-3]^N\frac{Q_{1}(u+2)}{Q_{1}(u-2)} - 
[u-1]^N\frac{Q_{1}(u+2)Q_{2}(u-3)} 
{Q_{1}(u-2)Q_{2}(u-1)} \\ 
\fl +[u-1]^N\frac{Q_{1}(u+2)Q_{2}(u-3)}{Q_{1}(u)Q_{2}(u-1)} 
-[u+1]^N\frac{Q_{1}(u+2)Q_{2}(u-3)}{Q_{1}(u)Q_{2}(u+1)}
+[u+1]^N\frac{Q_{2}(u-3)}{Q_{2}(u+1)}.
\nonumber 
\end{eqnarray}
We note that the function $\dot{{\cal T}}^{1}(u)$ associated with 
the contravariant Young superdiagram
 $\dot{\mu} =\phi; \dot{\lambda}=(1^1)$:
\begin{equation}
 \dot{{\cal T}}^{1}(u)=
    -\begin{array}{|c|}\hline 
     	-3 \\ \hline 
     \end{array}
   +
    \begin{array}{|c|}\hline 
     	-2 \\ \hline 
     \end{array}
   -\begin{array}{|c|}\hline 
     	-1 \\ \hline 
     \end{array} 
\end{equation}
coincides with Essler and Korepin's solution [EK] 
on supersymmetric $t-J$ model in the limit $q \to 1$ 
 except overall scalar factor after some redefinition
. 
 Pole freeness of the functions ${\cal T}^{a}(u)$ and 
$\dot{{\cal T}}^{a}(u)$ under the Bethe ansatz equation (\ref{BAE}) 
can be proved in the same way as {\it Theorem}\ref{polefree}.
\section{Example of non-distinguished simple roots case:
$p(1)=p(2)=1,p(3)=0$ grading}
\begin{figure}
    \setlength{\unitlength}{0.75pt}
    \begin{center}
    \begin{picture}(80,30) 
      \put(10,20){\circle{20}}
      \put(20,20){\line(1,0){40}}
      \put(70,20){\circle{20}}
      \put(62.929,12.9289){\line(1,1){14.14214}}
      \put(62.929,27.07107){\line(1,-1){14.14214}}
      \put(6,0){$\alpha_{1}$}
      \put(66,0){$\alpha_{2}$}
  \end{picture}
  \end{center}
  \caption{Dynkin diagram for the Lie superalgebra
  $sl(1|2)$ corresponding to the non-distinguished simple roots :
  ${\rm deg}(\alpha_1)=0, {\rm deg}(\alpha_2)=1$.}
  \label{dinkynno}
\end{figure}
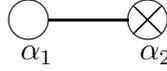

Let $\alpha_{1}$ and $\alpha_{2}$ be the simple roots of $sl(1|2)$ 
normalized so that 
 $(\alpha_{1}|\alpha_{1})=-2$,$(\alpha_{2}|\alpha_{2})=0$ and 
 $(\alpha_{1}|\alpha_{2})=(\alpha_{2}|\alpha_{1})=1$
  (see figure \ref{dinkynno}).
In this case, the sets (\ref{set}) and (\ref{set2}) become 
$J_{+}=\{ 3\}$, $J_{-}=\{1, 2\}$; 
$\dot{J}_{+}=\{ -3\}$, $\dot{J}_{-}=\{-1, -2\}$.
The function $z(a;u)=\framebox{$a$}_{u}$ $(a \in J)$ has the form  
\begin{equation}
\fl \framebox{1}=[u-2]^N \frac{Q_{1}(u+1)}{Q_{1}(u-1)}, 
\framebox{2}=[u]^N \frac{Q_{1}(u-3)Q_{2}(u)}{Q_{1}(u-1)Q_{2}(u-2)}, 
 \framebox{3}=[u]^N \frac{Q_{2}(u)}{Q_{2}(u-2)} 
\end{equation}
and the function $\dot{z}(a;u)=\framebox{$a$}_{u}$
 $(a \in \dot{J})$ has the form 
\begin{equation}
\fl  \framebox{-3}=[u+2]^N \frac{Q_{2}(u-1)}{Q_{2}(u+1)},
\framebox{-2}=[u]^N \frac{Q_{1}(u+2)Q_{2}(u-1)}{Q_{1}(u)Q_{2}(u+1)}, 
\framebox{-1}=[u]^N \frac{Q_{1}(u-2)}{Q_{1}(u)}.
\end{equation}
Here we assume the quantum spaces are labeled by Young superdiagrams 
with shapes  
$\tilde{\mu}=(1^1)$ and
 $\dot{\tilde{\mu}}=(1^1)$ respectively; for simplicity, 
 inhomogeneity parameters $w_i$ are set to $0$.
For example, for $\lambda=\phi; \mu =(2^1)$, (\ref{Tge1}) has the form
\begin{eqnarray}
  {\cal T}_{2}^{1}(u)=
    \begin{array}{|c|c|}\hline 
     	1 & 2 \\ \hline 
     \end{array}
   -
    \begin{array}{|c|c|}\hline 
     	1 & 3 \\ \hline 
     \end{array}
   -\begin{array}{|c|c|}\hline 
     	2 & 3 \\ \hline 
     \end{array}
   +\begin{array}{|c|c|}\hline 
     	3 & 3 \\ \hline 
     \end{array} \nonumber \\ 
\fl =
[u-3]^N [u+1]^N\frac{Q_{2}(u+1)}{Q_{2}(u-1)}-
[u-3]^N [u+1]^N\frac{Q_{1}(u)Q_{2}(u+1)}{Q_{1}(u-2)Q_{2}(u-1)}
 \nonumber \\ 
-[u-1]^N [u+1]^N\frac{Q_{1}(u-4)Q_{2}(u+1)}{Q_{1}(u-2)Q_{2}(u-3)} \\ 
 +[u-1]^N [u+1]^N\frac{Q_{2}(u+1)}{Q_{2}(u-3)} 
 \nonumber  
\end{eqnarray}
and for $\lambda =\phi; \mu=(1^2)$, (\ref{Tge1}) has the form 
\begin{eqnarray}
  {\cal T}_{1}^{2}(u)=
  \frac{1}{[u-1]^N}
  \left(
  \begin{array}{|c|}\hline 
     	1 \\ \hline 
     	1 \\\hline
     \end{array}
    +\begin{array}{|c|}\hline 
     	1 \\ \hline 
     	2 \\\hline
     \end{array}
  - 
    \begin{array}{|c|}\hline 
     	1 \\ \hline 
     	3 \\\hline
     \end{array}
  + 
  \begin{array}{|c|}\hline 
     	2 \\ \hline 
     	2 \\\hline
     \end{array}
  - 
  \begin{array}{|c|}\hline 
     	2 \\ \hline 
     	3 \\\hline
     \end{array} \right) \nonumber \\ 
  =
[u-3]^N\frac{Q_{1}(u+2)}{Q_{1}(u-2)} + 
[u-1]^N\frac{Q_{1}(u-4)Q_{1}(u+2)Q_{2}(u-1)} 
{Q_{1}(u-2)Q_{1}(u)Q_{2}(u-3)} \nonumber \\ 
\fl -[u-1]^N\frac{Q_{1}(u+2)Q_{2}(u-1)}{Q_{1}(u)Q_{2}(u-3)} 
  +[u+1]^N\frac{Q_{1}(u-4)Q_{2}(u+1)}{Q_{1}(u)Q_{2}(u-3)}\\
-[u+1]^N\frac{Q_{1}(u-2)Q_{2}(u+1)}{Q_{1}(u)Q_{2}(u-3)}.
\nonumber 
\end{eqnarray}
We note that the function $\dot{{\cal T}}^{1}(u)$ associated with 
the contravariant Young superdiagram with shape
 $\dot{\lambda} =\phi; \dot{\mu}=(1^1)$
\begin{equation}
 \dot{{\cal T}}^{1}(u)=
    \begin{array}{|c|}\hline 
     	-3 \\ \hline 
     \end{array}
   -
    \begin{array}{|c|}\hline 
     	-2 \\ \hline 
     \end{array}
   -\begin{array}{|c|}\hline 
     	-1 \\ \hline 
     \end{array} 
\end{equation}
coincides with Lai's solution [L] on supersymmetric $t-J$ model 
presented in [EK]  
in the limit $q \to 1$ 
except overall scalar factor after some redefinition 
.
Pole freeness of the functions ${\cal T}^{a}(u)$ and 
$\dot{{\cal T}}^{a}(u)$ under the Bethe ansatz equation (\ref{BAE}) 
can be proved in the same way as {\it Theorem}\ref{polefree}.
\section{Other representation of ${\cal T}^{a}$
and ${\cal T}_{m}$  }
For simplicity, we assume the vacuum part is formally trivial. 
Define the functions ${\cal A}^{a}$,
${\cal B}^{a}$,
${\cal A}_{m}$ and ${\cal B}_{m}$ 
 by the generating series such that  
\begin{eqnarray}
\fl  \sum_{k=-\infty}^{\infty} 
     {\cal A}_{k}(u+k-1)X^{k} 
     =(1-z(1;u)X)^{-1}\cdots (1-z(r+1;u)X)^{-1} \\ 
\fl  \sum_{l=-\infty}^{\infty} 
   {\cal B}^{l}(u+l-1)X^{l}  
   = (1-z(r+2;u)X)\cdots (1-z(r+s+2;u)X) 
\end{eqnarray} 
\begin{eqnarray}
\fl \sum_{k=-\infty}^{\infty} 
  {\cal B}_{k}(u+k-1)X^{k} 
  = (1+z(r+s+2;u)X)^{-1}\cdots (1+z(r+2;u)X)^{-1} \\ 
\fl \sum_{l=-\infty}^{\infty} 
 {\cal A}^{l}(u+l-1)X^{l}=
  (1+z(r+1;u)X)\cdots (1+z(1;u)X) 
\end{eqnarray} 
Combining these relations, we obtain
\begin{equation}
{\cal T}^{a}(u)=\sum_{l=0}^{min(r+1,a)}
  {\cal B}_{a-l}(u-l){\cal A}^{l}(u+a-l)
\end{equation}
\begin{equation}
{\cal T}_{m}(u)=\sum_{l=0}^{min(s+1,m)}
  {\cal A}_{m-l}(u-l){\cal B}^{l}(u+m-l).
\end{equation}
Note that these functions ${\cal A}_{m}(u)$ and ${\cal A}^{a}(u)$
are analogous to eigenvalue formulae of transfer matrices 
in dressed vacuum form 
of fusion $U_q(sl_{r+1}^{(1)})$ vertex model labeled by Young diagrams 
with shapes $(m^1)$ and $(1^a)$ respectively. 
We also note that the functions ${\cal B}^{a}(u)$ and 
${\cal B}_{m}(u)$ are analogous to
eigenvalue formulae of transfer matrices in dressed vacuum form 
of fusion $U_q(sl_{s+1}^{(1)})$ vertex model labeled by Young diagrams 
with shapes $(1^a)$ and $(m^1)$ respectively.
\section*{References}
\begin{harvard}
        \item[] [BB1]
        Balantekin A B and Bars I 1981 \JMP{\bf 22} 1149 
        
        \item[] [BB2]
        Balantekin A B and Bars I 1981 \JMP{\bf 22} 1810
        
        \item[] [BMR]
        Bars I, Morel B and Ruegg H 1983 \JMP{\bf 24} 2253
        
        \item[] [BR]
        Bazhanov V V and Reshetikhin N 1990 \JPA{\bf 23} 1477 
        
        \item[] [BS]
        Bazhanov V V and Shadrikov A G 1987 
        {\it Theor. Math. Phys.} {\bf 73} 1302
        
         \item[] [C]
        Cornwell J F 1989 {\it GROUP THEORY IN PHYSICS Vol 3
        Supersymmetries and Infinite-Dimensional Algebras}  
        (Academic press, New York)   
        
        \item[] [D]
        Drinfel'd V G 1988 {\it Sov.Math.Dokl} {\bf 36} 212
        
        \item[] [DM]
        Deguchi T and Martin P P 1992 {\it Int. J. Mod. Phys. A} {\bf 7} 
        {\it Suppl. 1A} 165
        
        \item[] [EK]
        Essler F H L and Korepin V E 1992 \PR {\it B} {\bf 46} 9147 
        
        \item[] [EKS1]
        Essler F H L, Korepin V E and Schoutens K 1992 \PRL 
        {\bf 68} 2960 
        
        \item[] [EKS2]
        Essler F H L, Korepin V E and Schoutens K 
        cond-mat/9211001; 1994 {\it Int. J. Mod. Phys.} {\bf B 8} 3205
        
        \item[] [FK]
        Foerster A and Karowski M 1993 \NP {\it B} {\bf 396} 611
        
        \item[] [H]
        Hirota R 1981 \JPSJ {\bf 50} 3787
        
        \item[] [Ka]
        Kac V 1977 {\it Adv. Math.} {\bf 26} 8  
        
         \item[] [KE]
        Korepin V E and Essler F H L eds. 1994 {\it Exactly solvable 
        models of strongly correlated electrons} 
        (World Scientific, Singapore)   
        
        \item[] [KLWZ] Krichever I, Lipan O, Wiegmann P and Zabrodin A  
         1997 {\it Commun. Math. Phys.} {\bf 188} 267  
        
        \item[] [K]
        Kuniba A  1994 \JPA{\bf 27} L113
        
        \item[] [KNH]
        Kuniba A, Nakamura S and Hirota R 1996 \JPA{\bf 29} 1759  
               
        \item[] [KNS1]
        Kuniba A, Nakanishi T and Suzuki J 1994 {\it Int. J. Mod. Phys.}
         {\bf A9} 5215 
          
        \item[] [KNS2]
        Kuniba A, Nakanishi T and Suzuki J 1994 {\it Int. J. Mod. Phys.}
         {\bf A9} 5267 
         
        \item[] [KOS]
        Kuniba A, Ohta Y and Suzuki J  1995 \JPA{\bf 28} 6211
        
        \item[] [KS1]
        Kuniba A and Suzuki J 1995 {\it Commun. Math. Phys.} {\bf 173} 225
        
        \item[] [KS2]
        Kuniba A and Suzuki J 1995 \JPA{\bf 28} 711  
 
        \item[] [Kul]
        Kulish P P 1986 {\it J. Sov. Math} {\bf 35} 2648
        
        \item[] [KulSk]
        Kulish P P and Sklyanin E K 1982 {\it J. Sov. Math} {\bf 19} 1596
        
        \item[] [L]
        Lai C K 1974 \JMP {\bf 15} 1675
        
        \item[] [M]
        Maassarani Z 1995 \JPA {\bf 28} 1305
        
        \item[] [MR]
        Martin P and Rittenberg V 1992 {\it Int. J. Mod. Phys. A} {\bf 7} 
        {\it Suppl. 1B} 707
        
        \item[] [N]
        Nazarov M L 1991 {\it Lett. Math. Phys} {\bf 21} 123
        
        \item[] [PF]
        Pfannm$\ddot{{\bf u}}$ller M P and Frahm H 1996 \NP {\bf B479} 575 
        
        \item[] [PS1]
        Perk J H H and Schultz CL in 1983 {\it Nonlinear Integrable Systems
        -Classical Theory and Quantum Theory} ed Jimbo M and Miwa T 
        (World Scientific, Singapore)
        
        \item[] [PS2]
        Perk J H H and Schultz CL 1981 \PL{\bf 84A} 407        
        
        \item[] [PT]
        Pragacz P and Thorup A 1992 {\it Adv. Math.} {\bf 95} 8  
        
        \item[] [R1]
        Reshetikhin N Yu 1983 {\it Sov. Phys. JETP} {\bf 57} 691 
        
         \item[] [R2]
        Reshetikhin N Yu 1987 {\it Lett. Math. Phys.} {\bf 14} 235 
        
        \item[] [RW]
        Reshetikhin N Yu and Wiegmann P B 1987 \PL{\bf B189} 125 
        
        \item[] [Sc]
        Schultz C L 1983 {\it Physica}{\bf A122} 71
        
        \item[] [S1]
        Suzuki J 1992 \JPA{\bf 25} 1769
        
        \item[] [S2]
        Suzuki J 1994 \PL{\bf A195} 190
        
        \item[] [Su]
        Sutherland B 1975 {\it Phys. Rev. } {\bf 12B} 3795
        
        \item[] [T]
        Tsuboi Z {\it solv-int}/9610011; 1997 \JPSJ {\bf 66} 3391
        
        \item[] [TK]
        Tsuboi Z and Kuniba A  1996 \JPA{\bf29} 7785
        
        \item[] [Y]
        Yamane H preprent {\it q-alg}/9603015 
        
        \item[] [ZB]
        Zhou Y K and Batchelor M T 1997 {\it Nucl. Phys.} 
        {\bf 490B} 576 
                        
\end{harvard}
\end{document}